\documentclass[epj]{webofc}
\usepackage[varg]{txfonts}   
\usepackage{cleveref}
\usepackage{pifont}
\usepackage{natbib}
\usepackage{subfigure,epsfig,epstopdf}
\usepackage{bm}
\usepackage{enumerate}

\newcommand{\be}{\begin{equation}}
\newcommand{\ee}{\end{equation}}
\newcommand{\bea}{\begin{eqnarray}}
\newcommand{\eea}{\end{eqnarray}}

\def\be{\begin{equation}}
 \def\ee{\end{equation}}
 \def\l{\lambda}
 \def\a{\alpha}

 \def\m{\mu}
 \def\n{\nu}

 \def\ds#1{#1\kern-1ex\hbox{/}}
 \def\sla{\raise.15ex\hbox{$/$}\kern-.57em}

 \def\({\left(}
 \def\){\right)}
 \def\[{\left[}
 \def\]{\right]}
\woctitle{ICNFP 2015}
\begin{document}
\title{Aspects of Meson Condensation}
%
%

\author{Andrea Mammarella \inst{1}\fnsep\thanks{\email{andrea.mammarella@lngs.infn.it}} 
}

\institute{Laboratori Nazionali del Gran Sasso 
          }

\abstract{
In this work pion and kaon condensation in the framework of chiral perturbation theory is studied. I consider a system at vanishing temperature with nonzero isospin chemical potential and strangeness chemical potential; meson masses and mixing in the normal phase, the pion condensation phase and the kaon condensation phase are described. There are differences with previous works, but the results presented here are supported by both theory group analysis and by direct calculations. Some pion decay channels in the normal and the pion condensation phases are studied, finding a nonmonotonic behavior of the $\Gamma$ decay as a function of $\mu_I$.
}
\maketitle
\section{Introduction}
\label{intro}
The properties of strongly interacting matter in an isospin and/or strangeness rich medium are relevant  in a wide range of phenomena including  the astrophysics of compact stars and heavy-ion collisions.  It is known that depending on the value of the isospin chemical potential, $\mu_I$, and on the value of the strangeness chemical potential,  $\mu_S$, three different phase can be realized: the normal phase, the pion condensed ($\pi c$) phase and the kaon condensed ($Kc$) phase~\cite{Migdal:1990vm,Son:2000xc,Kogut:2001id}. The realization of a mesonic condensate can drastically change the low energy properties of matter, including the mass spectrum and the lifetime of mesons.\\

Previous analyses of the meson condensed phases by QCD-like theories were developed in \cite{Kogut:1999iv, Kogut:2000ek}. Pion condensation in two-flavor quark matter was studied in~\cite{Son:2000xc, Son:2000by} and in three-flavor quark matter in~\cite{Kogut:2001id}. In particular,  the phase diagram as a function of $\mu_I$ and $\mu_S$ was presented in~\cite{Kogut:2001id}. Finite temperature effects in  $SU(2)_L \times SU(2)_R$ chiral perturbation theory ($\chi$PT) have been studied in~\cite{Loewe:2002tw, Loewe:2004mu, He:2005nk, Xia:2014bla}. One remarkable property of quark matter with nonvanishing isospin chemical potential is that it is characterized by a real measure, thus the lattice realization can be performed with standard numerical algorithms~\cite{Alford:1998sd, Kogut:2002zg}. The $\pi c$ phase and the $K c$ phase have  been studied by NJL models in~\cite{Toublan:2003tt, Barducci:2004tt, Barducci:2004nc} and by random matrix models in \cite{Klein:2004hv}. All these models find results in qualitative and quantitative agreement, and in particular, the phase diagram of matter has been firmly established. However, regarding  the low energy mass spectrum in three-flavor quark matter, we found that it was only studied in~\cite{Kogut:2001id}. Our results are in disagreement with those of~\cite{Kogut:2001id}, the most relevant difference is in the  mixing between mesonic states.  Regarding the pion decay, previous works focused on density and temperature effects in standard decay channels~\cite{Barducci:1990sv, Dominguez:1993kr,Loewe:2011tm}, but not all the decay channels have been considered. \\

In this article I briefly review how to include chemical potentials in $\chi PT$ ~\cite{Gasser:1983yg, Leutwyler:1993iq, Ecker:1994gg,  Scherer:2002tk, Scherer:2005ri}, then I describe some phenomenological properties related to this inclusion, like the existence of the different phases already listed, the meson masses and mixing. I will also show how the inclusion of chemical potentials affects pion decay channels \cite{Mammarella:2015pxa}.

The paper is organized as follows. In Sec.~\ref{mod} I describe the model and how it predicts different phases. In Sec.~\ref{mm} I show how group theory tools can be used to calculate the mass eigenstates in the condensed phases and I list and discuss the results obtained. In Sec.\ref{pion} I discuss the impact of chemical potentials on charged pion decays. In Sec. ~\ref{conc} I summarize the results.

\section{Model} \label{mod}
\subsection{Lagrangian and definitions}
In this section I briefly review the  model that I am going to use in the following. It is the one described in \cite{Kogut:2001id}.
The general ${\cal O}(p^2)$ Lorentz invariant Lagrangian density describing the pseudoscalar mesons can be written as 
\be\label{eq:Lagrangian_general}
{\cal L} = \frac{F_0^2}{4} \text{Tr} (D_\nu \Sigma D^\nu \Sigma^\dagger) + \frac{F_0^2}{4} \text{Tr} (X \Sigma^\dagger + \Sigma X^\dagger )\,,
\ee
where $\Sigma$ corresponds to the meson fields, $X=2 B_0( s + i p)$ describes scalar and pseudoscalar external fields and the covariant derivative is defined as 
\be
D_\mu \Sigma = \partial_\mu\Sigma - \frac{i}{2} [v_\mu,\Sigma] - \frac{i}{2} \{a_\mu,\Sigma \}\,,
\ee
with  $v_\mu$ and  $a_\mu$   the external vectorial and axial currents, respectively. The  Lagrangian has two free parameters $F_0$ and $B_0$, related to the pion decay and to the quark-antiquark condensate, respectively, see for example~\cite{Gasser:1983yg, Leutwyler:1993iq, Ecker:1994gg,  Scherer:2002tk, Scherer:2005ri}. 

The Lagrangian density is invariant under $SU(N_f)_L \times SU(N_f)_R $  provided the meson field transform as 
\be\label{eq:transSigma}
\Sigma \to R \Sigma L^\dagger\,,
\ee
and the chiral symmetry breaking corresponds to the spontaneous global symmetry breaking $SU(N_f)_L \times SU(N_f)_R \to SU(N_f)_{L+R}  $. In standard $\chi$PT, the mass eigenstates are charge eigenstates as well. Thus  mesons are particles with a well defined mass  and charge. The presence of a medium can change this picture. In particular, if the vacuum carries  an electric charge, then the mass eigenstates will not typically be charge eigenstates. The presence of a medium can be taken into account by considering appropriate external currents in Eq.~\eqref{eq:Lagrangian_general}. 

At vanishing temperature the vacuum is determined by maximizing  the Lagrangian density with respect to the external currents. 
The pseudoscalar mesons are then described as oscillations around the vacuum. We use the same nonlinear representation of~\cite{Kogut:2001id} corresponding to 
\be\label{eq:sigma}
\Sigma= u \bar \Sigma u \qquad \text{with} \qquad u=e^{i  T \cdot \phi/2} \,,
\ee
where $T_a$ are the $SU(N_f)$ generators and  $ \bar \Sigma $ is a generic $SU(N_f)$ matrix to be determined by maximizing the static Lagrangian. The reasoning behind the  above expression  is that under $SU(N_f)_L \times SU(N_f)_R $ mesons can be identified as   the fluctuations of the vacuum as in Eq.~\eqref{eq:transSigma} with $\theta_a^R = - \theta_a^L = \phi_a $. 

In the following we will assume that $a_\mu=0$, $p=0$, $X=2 G M$, where $M$ is the $N_f \times N_f$ diagonal quark mass matrix and $G$ is a constant, that with these conventions is equal to $B_0$. 
Moreover, we will assume that 
\be \label{ext}
v^\nu = -2 e Q A^\nu - 2 \mu \delta^{\nu 0} = - \frac{2}3 (\mu_B-\mu_S) I \delta^{\nu 0}- \tilde A^\nu_3 \lambda_3 - \tilde A^\nu_8 \lambda_8\,,
\ee
 meaning that the vectorial current consists of the electromagnetic field and a quark chemical potential, with $\mu$  a $SU(N_f) \times SU(N_f)$  matrix in flavor space. Its explicit expression is:

\begin{align}
\mu=\text{diag}\left(\mu_u,\mu_d,\mu_s\right)= \text{diag}\left(\frac13 \mu_B+\frac12 \mu_I,\frac13 \mu_B-\frac12 \mu_I, \frac13
\mu_B-\mu_S\right)=\frac{\mu_B-\mu_S}3 I + \frac{\mu_I}2 \lambda_3 + \frac{\mu_S}{\sqrt{3}} \lambda_8\,,
\end{align}
It is important to remark that this model only holds for $|\m_B|\lesssim 940$ MeV, $|\m_I|\lesssim 770$ MeV and $|\m_S|\lesssim 550$ MeV \cite{Kogut:2001id}.

\subsection{Groud state and different phases} \label{ground}
To find the ground state we have to substitute (\ref{eq:sigma}) in the Lagrangian (\ref{eq:Lagrangian_general}). It is not necessary to use a complete SU(3) parametrization for $\bar{\Sigma}$, but is sufficient:
\begin{eqnarray}
\bar \Sigma=\left( \begin{array}{ccc} 1 & 0 & 0 \\
                                 0 & \cos \beta & -\sin \beta \\
                                 0 & \sin \beta & \cos \beta
                   \end{array} \right) \;
            \left( \begin{array}{ccc} \cos \alpha & \sin \alpha  & 0 \\
                                 -\sin \alpha & \cos \alpha & 0 \\
                                 0 & 0 & 1
                   \end{array} \right) \;
            \left( \begin{array}{ccc} 1 & 0 & 0 \\
                                 0 & \cos \beta & \sin \beta \\
                                 0 & -\sin \beta & \cos \beta
                   \end{array} \right),
\label{SadP}
\end{eqnarray}
because $\bar{\Sigma}$ has to be orthogonal to the chemical potential in the SU(3) generator space \cite{Mammarella:2015pxa}.\\
Substituting (\ref{SadP}) in (\ref{eq:Lagrangian_general}) and maximizing, we find three different vacua, that implies that there are three ground states and therefore three different phases \cite{Kogut:2001id}:
\begin{itemize}
\item Normal phase: \begin{align} \mu_I&<m_\pi\,,\\ \mu_S&<m_K-\frac12 \mu_I\,,\end{align}
 characterized by
\be
     \alpha_N=0,  \quad\beta_N \in (0,\pi), \quad \bar{\Sigma}_N=\text{diag}(1,1,1)\,.
\ee
\item Pion condensation phase: \begin{align} \mu_I&>m_\pi\,, \\ \mu_S&<
\frac{-m_\pi^2+\sqrt{(m_\pi^2-\mu_I^2)^2+4 m_K^2 \mu_I^2} }{2 \mu_I}\,,
\end{align}
 characterized by
\be
    \cos \alpha_{\pi}=\left(\frac{m_\pi}{\mu_I}\right)^2,  \quad \beta_{\pi} =0\,,\ee  \begin{align} \bar{\Sigma}_{\pi}&=\left( \begin{array}{ccc}
                            \cos \alpha_\pi & \sin \alpha_\pi & 0 \\
                            -\sin \alpha_\pi & \cos \alpha_\pi & 0\\
                            0 & 0 & 1
                            \end{array} \right) \\ &= \frac{1+2 \cos\alpha_\pi}{3} I + i \lambda_2 \sin\alpha_\pi+ \frac{\cos\alpha_\pi-1}{\sqrt{3}}\lambda_8\,. \nonumber
                            \label{pion}
\end{align}
\item Kaon condensation phase: 
\begin{align}
\mu_S&>m_K-\frac12 \mu_I\,, \\ 
\mu_S&>
\frac{-m_\pi^2+\sqrt{(m_\pi^2-\mu_I^2)^2+4 m_K^2 \mu_I^2 }}{2 \mu_I}\,,
\end{align}
characterized by
\be
\cos \alpha_K=\left( \frac{m_K}{\frac12 \mu_I+\mu_S}\right)^2\,,  \beta_K=\pi/2\,,\ee 
\begin{align} \label{kaon} \bar{\Sigma}_K
=&\left( \begin{array}{ccc}
                            \cos \alpha & 0 & \sin \alpha  \\
                            0 & 1 & 0\\
                            -\sin \alpha & 0 & \cos \alpha
                            \end{array} \right)=\\
&\frac{1+2 \cos\alpha_K}{3} I+ \frac{\cos\alpha_K-1}{2\sqrt{3}}\left(\sqrt{3}\lambda_3-\lambda_8\right) + i \lambda_5 \sin\alpha_K\,. \nonumber 
\end{align}
Note that the kaon condensation  can only happen for
\be \mu_S > \bar\mu_S=m_K - \frac{m_\pi}2\,.\ee 

\end{itemize}

\section{Meson Masses and Mixing} \label{mm}
\subsection{Mixing}
As we have seen in Sec. \ref{ground} the ground state in the condensed phases is not diagonal and thus has SU(3) charges that cause symmetry breaking. It is useful to study the breaking pattern to learn something about the possible meson mixing in the condensed phases.

The starting Lagrangian has an $SU(3)_L \times SU(3)_R$ simmetry, broken to $SU(3)_V$ by the quark masses. The introduction of chemical potentials via the external current (\ref{ext}) further reduce this symmetry to $U(1)_{L+R}\times U(1)_{L+R}$, meaning that the $\l_3$ and $\l_8$ terms in (\ref{ext}) break isospin and hypercharge conservation. When the system enters one of the condensed phases the vacuum acquires a charge and thus there is no symmetry left.

\begin{figure}[th!]
\centering
\includegraphics[width=6.cm]{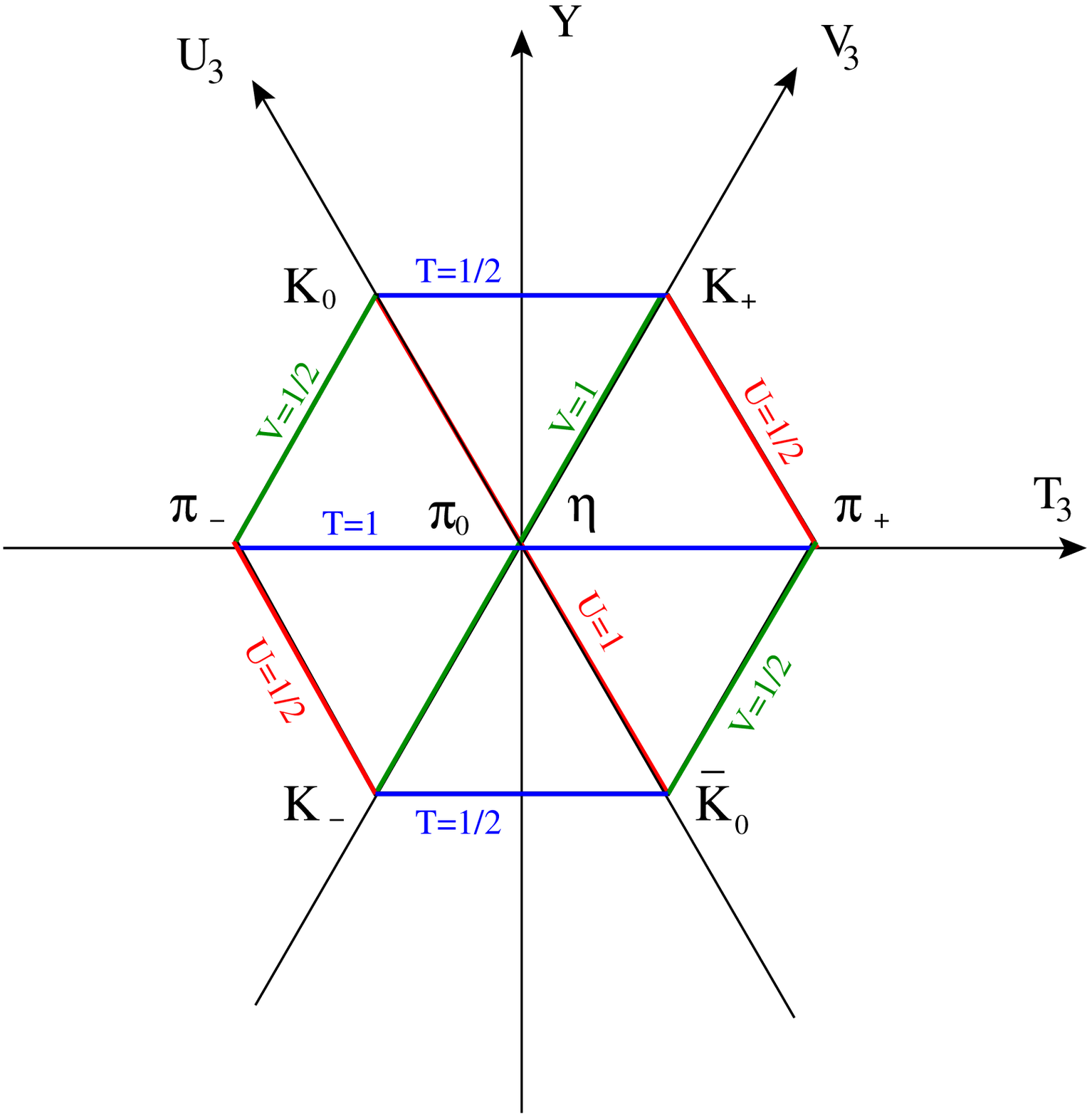}
\includegraphics[width=6.cm]{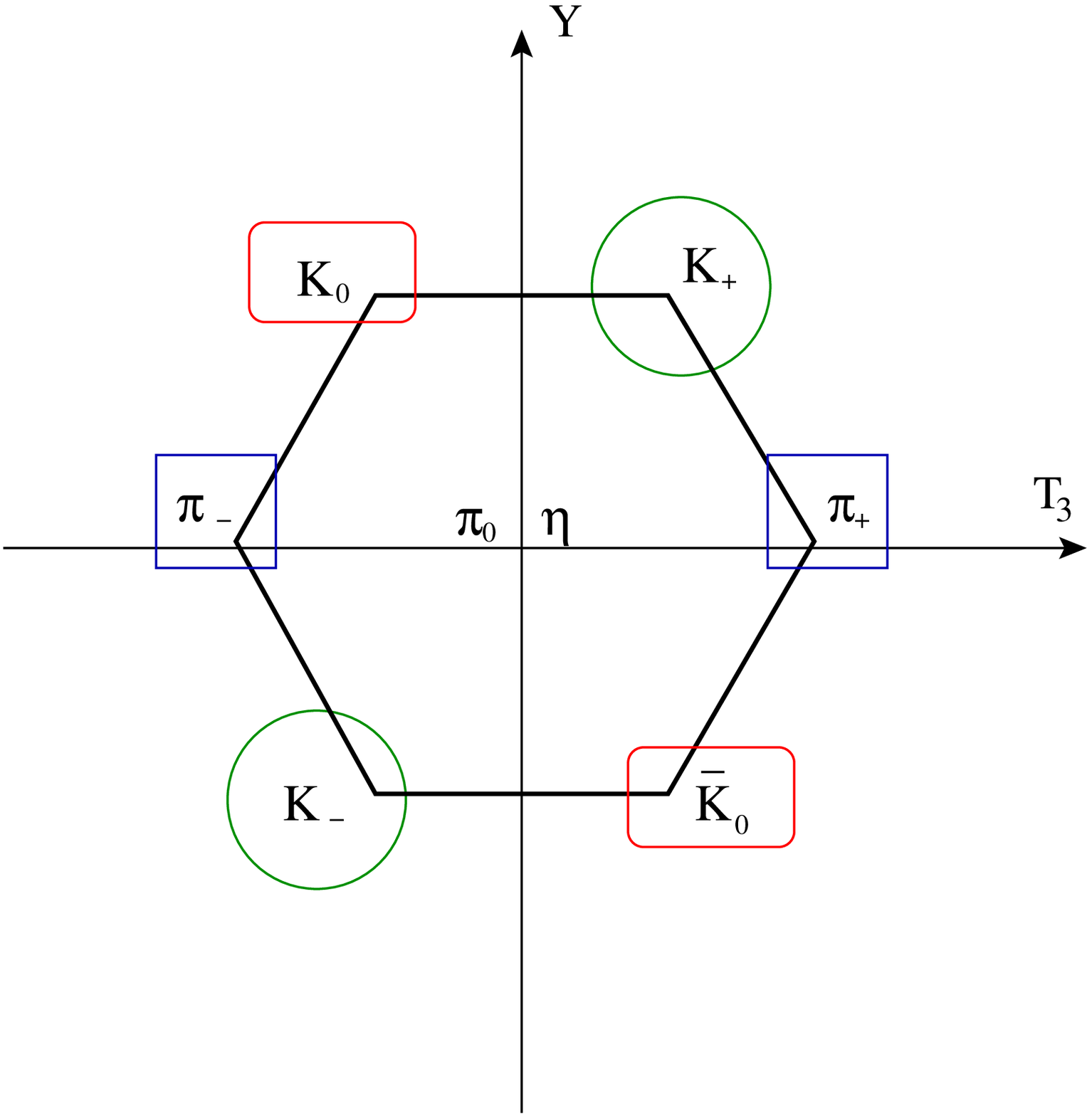}
\caption{(color online) Weight diagrams of the mesonic octet. In the top panel we have reported the axes corresponding to the third component of isospin (or $T$-spin), $U$-spin, $V$-spin and hypercharge, $Y$. Any mesonic state belongs to a multiplet of   $T$-spin, $U$-spin and $V$-spin as indicated in the diagram. In the bottom panel we have indicated the mesonic states having the same $T$-spin, $U$-spin and $V$-spin quantum numbers. Only  the states marked with the same symbol can mix.   The $\pi_0$ and the $\eta$  are not simultaneous    $T$-spin, $U$-spin and $V$-spin eigenstates; their mixing depends on the spontaneously induced charge of the vacuum. }
\label{fig:mixing}
\end{figure}

In all these cases we can use the quantum numbers of the $SU(2)$ subgroups of $SU(3)$ to label the states and to find out the ones that can mix. In figure \ref{fig:mixing} these quantum numbers are represented. We need only two of them, because they are not independent. In table \ref{table:mixing} are shown the states that can mix and the related quantum numbers.

\begin{table}[h!]
\begin{center}
\begin{tabular}{|c|c|}\hline
Mixing states & $(T,U)$\\ \hline
$\pi_+,\pi_-$ & $(1,1/2)$ \\ \hline
$K_+,K_-$ & $(1/2,1/2)$ \\ \hline
$K_0,\bar K_0$ & $(1/2,1)$ \\
\hline
\end{tabular}
\end{center}
\caption{Mixing mesons  with the corresponding $T$-spin and $U$-spin quantum numbers. These quantum numbers label the $SU(3)$ subspace spanned by the corresponding mesonic states.  The $\pi_0$ and the $\eta$ do not appear because they are not $U$-spin eigenstates.}
\label{table:mixing}
\end{table}%

Unfortunately, this is not sufficient to determine if $\pi_0$ and $\eta$ can mix, because they do not have a well defined U and V spin, so we have to study deeply how the ground state affect them.
Let us first consider the normal phase. In the normal phase there is no operator that can induce the mixing of the mesonic  states, thus the mesonic states remain unchanged but the $Q_3$ and $Q_8$ charges will induce Zeeman-like mass splittings.

In any of the condensed phases, there is an additional charge that is spontaneously induced, and the corresponding operator will lead to mixing.  

Let us first focus on isospin (or $T$-spin). We have to consider two cases.  Suppose that the vacuum has a charge that  commutes with $T^2$, as in the $\pi c$ phase,  say the charge corresponding to $T_2 = i( T_- - T_+)$, see Eq.~\eqref{pion}.  The $T_\pm$ operators can induce mixing among the charged pions and among the kaons. On the other hand, $T$-spin conservation does not allow the $\vert\pi_0\rangle =  \vert T=1, T_3=0\rangle$  to mix with the $\vert\eta\rangle = \vert T=0, T_3=0\rangle$.  

Now suppose instead that the vacuum has a charge that  does not commute with $T^2$ as in the $Kc$ phase, see Eq.~\eqref{kaon}. Any operator that does not commute with isospin will commute with $U$-spin or with $V$-spin. In the $K c$ phase  $Q_5 \vert 0 \rangle \neq 0$, then the vacuum is not invariant under this charge. However, since  $[T_5,U]=0$  it follows that $U$-spin is conserved. The lowering and raising operators inducing the mixing will be $U_\pm$. Regarding the $\pi_0$ and the $\eta$, in this case  we have that $\vert U=1, U_3=0\rangle$ and $\vert U=0, U_3=0\rangle$ do not mix. Since $\vert U=1, U_3=0\rangle = \frac{\vert\pi_0\rangle + \sqrt{3}\vert \eta\rangle}{2}$ and $\vert U=0, U_3=0\rangle = \frac{\sqrt{3}\vert  \pi_0\rangle -\vert \eta\rangle}{2}$, these will be the mass eigenstates.   

\subsection{Masses}
\begin{figure}[th!]
\includegraphics[width=8.cm]{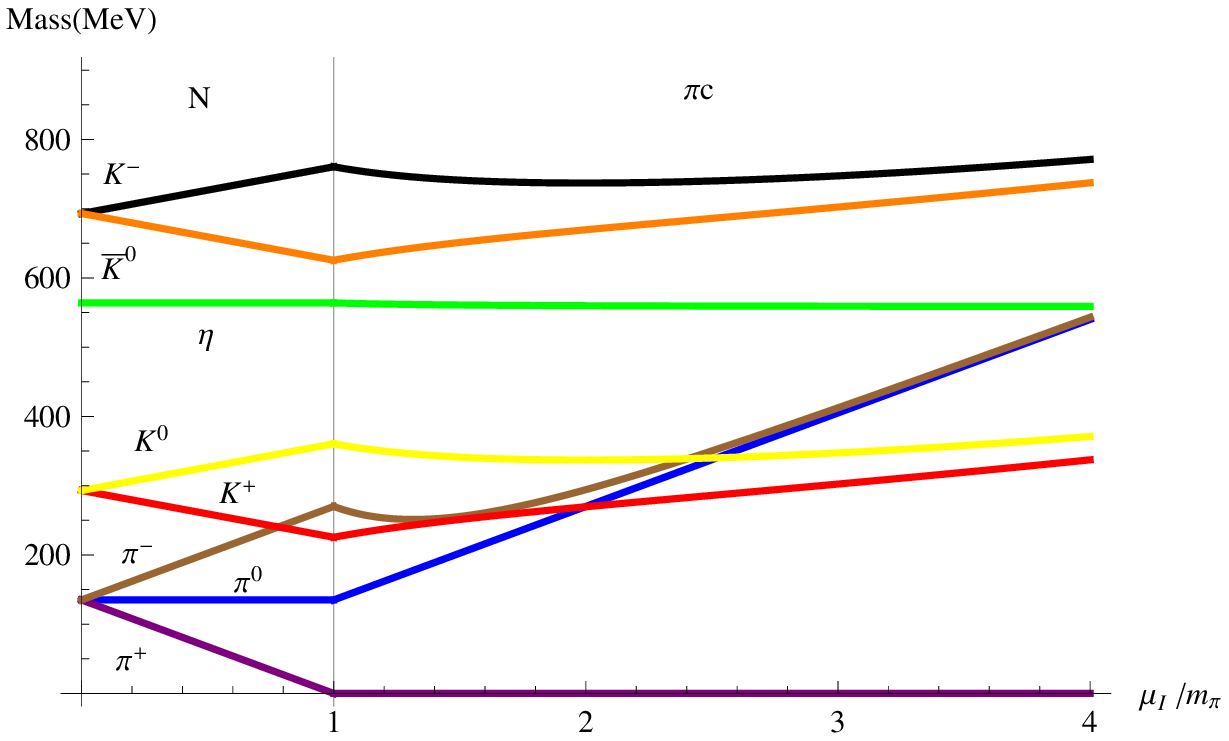}
\includegraphics[width=8.cm]{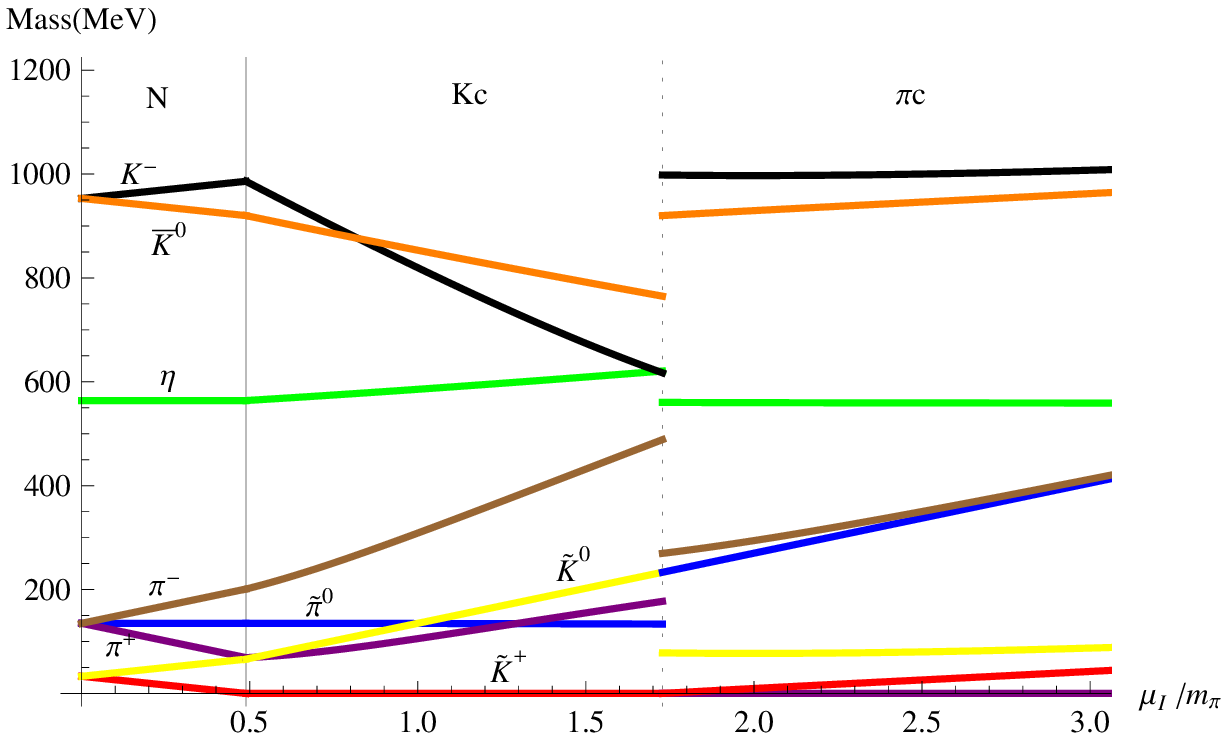}
\includegraphics[width=8.cm]{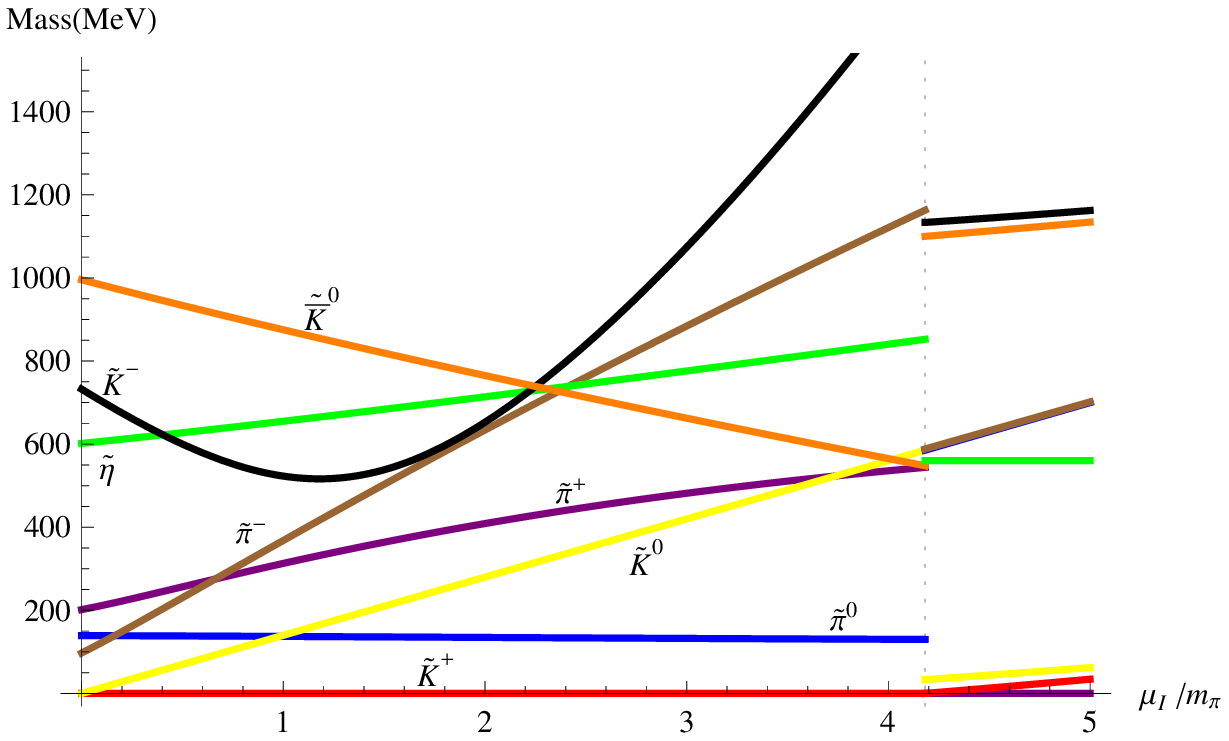}
\caption{(color online) Mass spectrum of the pseudoscalar mesonic octet.   Top panel, results obtained  for $\mu_S=200$ MeV. The vertical solid line represents the second order phase transition between the normal phase and the pion condensation phase.  In this case the strange quark chemical potential is below the threshold value for kaon condensation, $425$ MeV, thus the kaon condensed phase does not take place for any value of $\mu_I$.  Middle panel, results obtained  for  $\mu_S=460$ MeV. The vertical solid line represents the second order phase transition from the normal phase to the kaon condensation phase. The dashed line corresponds to the first order phase transition between the kaon condensed phase to the pion condensed phase. Bottom panel, results obtained  for  $\mu_S=550$ MeV, corresponding to the largest value of $\mu_S$.}
\label{fig:masses}
\end{figure}

The mass eigenstates are found diagonalizing the Lagrangian in the different phases. They present the mixing predicted by the group theory analysis of the previous subsection. Remember the definitions:
\begin{align}
  m_\pi^2&= 2 G m/F_0^2\,,  \\
  m_K^2&= G (m+m_s)/F_0^2\,, \\
  m_{\eta^0}^2&= 2 G (m+2 m_s)/3 F_0^2=(4 m_K^2-m_\pi^2)/3\,.
\end{align}
Meson masses have been calculated in \cite{Mammarella:2015pxa}. They are, in the normal phase:\begin{align}
    m_{\pi^0}&=m_\pi\,, \\
    m_{\pi^\pm}&=m_\pi \mp \mu_I\,,\label{eq:pi+-masses}\\
    m_{\eta^0}&=    \sqrt{(4 m_K^2-m_\pi^2)/3}\,, \\
     m_{K^\pm}&=m_K\mp\frac12 \mu_I\mp\mu_S\,,  \\
    m_{K^0/\bar K^0}&=m_K\pm\frac12 \mu_I\mp\mu_S \,.
\label{eq:norm_masses}
\end{align} 
In the condensed phases, as said, the mass eigenstates are mixture of normal phase mesons, so I will list them with a $\sim$. They have been calculated in \cite{Mammarella:2015pxa}. In the pion condensation phase, their masses are:
\begin{align}
m_{\tilde{\pi_0}}&=\m_I \\
m_{\tilde{\pi_+}}&=0 \\
m_{\tilde{\pi_-}}&=\frac{\sqrt{3 m_\pi^4 +\mu_I^4}}{\mu_I}\\
m_{\tilde{\eta}}^2&=m_{\eta}^2+\frac{1}{3} m_{\pi}^2\(\frac{m_\pi^2-\mu_I^2}{\mu_I^2}\)\, \\
m_{\tilde{K}^-/\tilde{K}^+}&=\pm\frac{1}{2}\(\frac{m_\pi^2}{\mu_I}+2\mu_S\pm\sqrt{\(\frac{m_\pi^2}{\mu_I}+2\mu_S  \)^2+4 M_4^2}  \)  \\
m_{\tilde{K}^0/\tilde{\bar{K}}^0}&=\pm\frac{1}{2}\(\frac{m_\pi^2}{\mu_I}-2\mu_S\pm\sqrt{\(\frac{m_\pi^2}{\mu_I}-2\mu_S  \)^2+4 M_6^2}  \)\,.  \\
\end{align} with:
\begin{align}
s_{45}&=\mu_I\cos\alpha +2\mu_S\,,  \\
M_4^2&=M_5^2=m_k^2+\frac{1}{4}\mu_I^2-\mu_S^2-\frac{1}{2}(\mu_I^2+2\mu_I\mu_S)\cos\alpha_\pi\,,\\
s_{67}&=\mu_I \cos \alpha_\pi-2\mu_S,  \\
M_6^2&=M_7^2=m_k^2+\frac{\mu_I^2}{4}-\mu_S^2-\frac{\mu_I^2-2\mu_I\mu_S}{2}\cos \a_\pi\,.
\end{align}
In the kaon condensation phase, their masses are:
\begin{align}
m_{\tilde\pi^\pm}&= 
\mp\frac{1}{2}\(\frac{\mu_I}{2}(3+\cos \a_K)+\mu_S(\cos \a_K-1)\mp\sqrt{\(\frac{\mu_I}{2}(3+\cos \a_K)+\mu_S(\cos \a_K-1)  \)^2+4 M_1^{'2}} \)\,, \\
m_{\tilde{K}^+}&=0\,,    \\
m_{\tilde{K}^-}&= \sqrt{M_5^{'2}+u_{45}^2}\,, \\
m_{\tilde{\bar{K}}^0/\tilde K^0}&=\pm\frac{1}{2} (u_{67}\pm\sqrt{u_{67}^2+4M_6^{'2}})\,, \\
m_{\tilde\pi^0/\tilde\eta}&=\sqrt{\frac{M_3^{'2}+M_8^{'2}\mp\sqrt{4 u_{38}^2+(M_3^{'2}-M_8^{'2})^2}}{2}}\,.
\end{align}
with:
\begin{align}
u_{12}&=\frac{1}{2}((3+\cos \a_K)\mu_I+2(\cos \a_K-1)\mu_S)\,,    \\
M_1^{'2}&=M_2^{'2}=m_\pi^2-\frac{1}{2}\mu_I^2(1+\cos \a_K)-4\mu_I\mu_S(\cos \a_K-1)\,,   \\
M_5^{'2}&= m_k^2 \cos \a_K-\frac{1}{4}(\mu_I+2\mu_S)^2  \cos (2\a_K)\,,  \\
u_{45}&= \frac{1}{4} \( \frac{1+3\cos \a_K)}{\cos \a_K}\) (\mu_I+2\mu_S)\,,  \\
M_6^{'2}&=M_7^{'2}= m_k^2+\frac{\mu_I-2\mu_S}{4}((\cos\a(\mu_I+2\mu_S)-2\mu_I)\,,   \\
u_{67}&=\frac{1}{2}((-3+\cos\a)\mu_I+2(1+\cos\a)\mu_S)\,, \\
M_3^{'2}&=\frac{1}{24}\left[ \cos \alpha_K\left(2  m_k^2 +9  m_\pi^2+6 \frac{G}{F_0^2}m_s\right)\right.\nonumber\\ &\left.+m_\pi^2 (16 -6   \cos^2 \a_K)\right] \,,  \\
M_8^{'2}&=\frac{1}{8 \cos \a_K}(2 m_k^2+(1+2 \cos\a_K)m_\pi^2\nonumber\\ &+6\frac{G}{F_0^2}m_s \cos \a_K)\,, \\
u_{38}&=\frac{1}{8 \sqrt{3}}(-2m_k^2+(3+2\cos\a_K)m_\pi^2-6 \frac{G}{F_0^2} m_s)\,.
\end{align}
Plot of these masses as a function of $\m_I$ for fixed values of $\m_S$ are shown in figure \ref{fig:masses}.

\section{Pion decays} \label{pion}
\begin{figure}[t]
\centering
\includegraphics[width=7.cm]{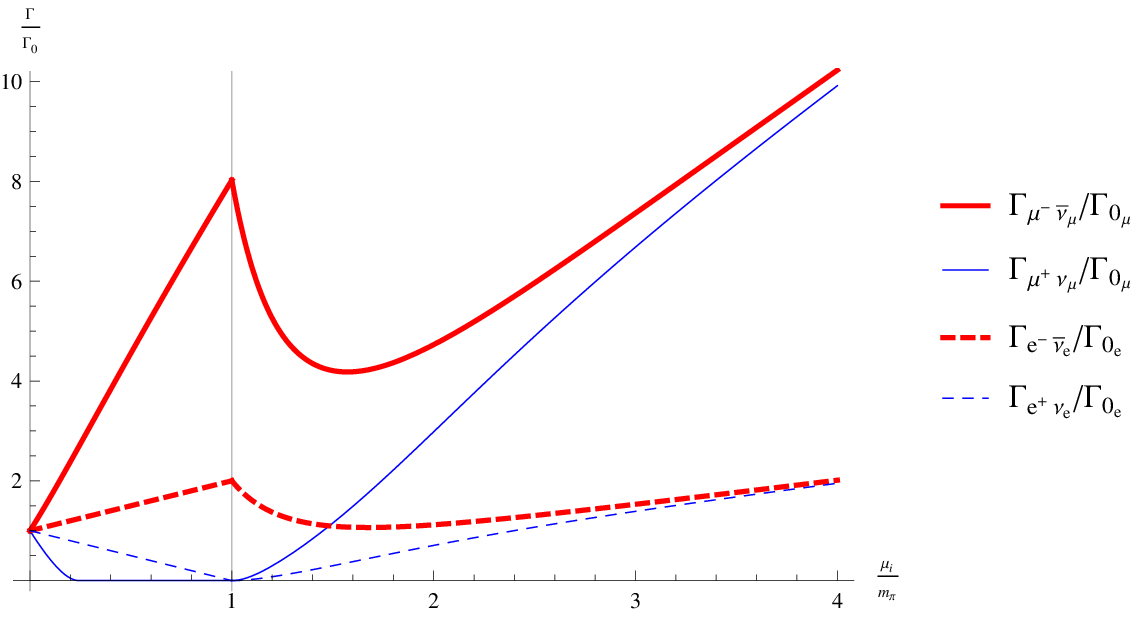}
\includegraphics[width=7.cm]{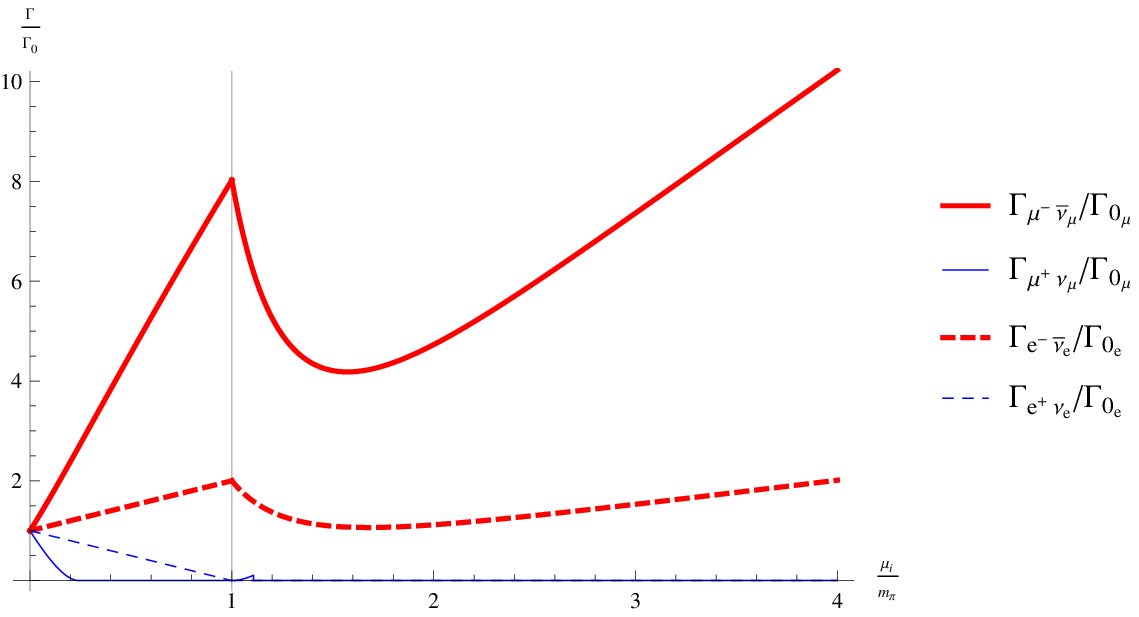}
\caption{(color online) Normalized leptonic decay rates of charged pions in normal phase  and in the  condensed phase normalized to the value in vacuum. The phase transition between the normal phase and the pion condesed phase corresponds to the  solid vertical line. Top, results obtained for vanishing leptonic chemical potential. Bottom, results obtained assuming weak equilibriub. In this case the decay in  positively charged leptons is Pauli blocked. In both plots the thick solid line represents $\Gamma_{\m^-\bar{\n}_\m}/\Gamma_{0_\m}$, the thin solid line represents  $\Gamma_{\m^+\n_\m}/\Gamma_{0_\m}$, the thick dashed line represents  $\Gamma_{e^-\bar{\n}_e}/\Gamma_{0_e}$ and the thin dashed line represents  $\Gamma_{e^+\n_e}/\Gamma_{0_e}$}
\label{fig:gammas}
\end{figure}

It has been shown how the introduction of chemical potentials can change meson masses and mixing. Here I will describe how it affects the charged pions decay. The processes are:
\begin{align}
\pi^+ &\rightarrow \ell^+ \nu_\ell\,,  \\
\pi^- &\rightarrow \ell^- \bar{\nu}_\ell\,,
\end{align}
that in the Standard Model (SM) have the decay width:
\be
\Gamma^0_{\pi \rightarrow \ell \nu_\ell}=\frac{G_F^2 F_0^2 V_{ud}^2 m_\ell^2 m_{\pi}}{4 \pi} \( 1-\frac{m_\ell^2}{m_{\pi}^2} \)^2 ,
\ee where $G_F$ is the Fermi constant, $V_{ud}$ is the $ud$ CKM matrix element, $m_\ell$ and $m_\pi$ are the lepton and pion masses.\\
For the following calculation I will assume $\m_S=0$ for simplicity.\\
In the normal phase chemical potentials do not mix states, so the only change is that we have to replace $m_\pi$ with $m_{\pi_\pm}$ from (\ref{eq:pi+-masses}).\\
In the pion condensation phase the $\tilde{\pi}_+$ is massless, so it does not decay, while the $\tilde{\pi}_-$ is a mixture of $\pi_\pm$ so it can decay in both 
$\ell^+\n_\ell$ and $\ell^-\bar{\n}_\ell$. The related decay widths are:
\begin{align}
\frac{\Gamma_{\tilde{\pi}_-\rightarrow \ell^+\nu_\ell}}{\Gamma^0_{\pi \rightarrow \ell \nu_\ell}}&=\frac{|U_{21}^* \cos \a+i U^*_{22}|^2}{2}\frac{m_{\tilde{\pi}^-}}{m_\pi} \left(\frac{1-m_\ell^2/m_{\tilde\pi^-}^2}{1-m_\ell^2/m_{\pi}^2}\right)^2\,,\\
\frac{\Gamma_{\tilde{\pi}_-\rightarrow \ell^-\bar\nu_\ell}}{\Gamma^0_{\pi \rightarrow \ell \nu_\ell}}&=\frac{|U_{21}^* \cos \a-i U^*_{22}|^2}{2}\frac{m_{\tilde{\pi}^-}}{m_\pi} \left(\frac{1-m_\ell^2/m_{\tilde\pi^-}^2}{1-m_\ell^2/m_{\pi}^2}\right)^2\,,
\end{align}
where:
\be
\left( \begin{array}{c}  \tilde\pi_+ \\ \tilde\pi_- \end{array} \right) = \left(\begin{array}{cc} U_{11} & U_{12} \\ U_{21} & U_{22}\end{array}  \right ) \left( \begin{array}{c}  \pi_+ \\ \pi_- \end{array} \right)\,, 
\label{eq:inverted}
\ee
The plot of these decay widths are shown in figure \ref{fig:gammas}.

\section{Conclusion} \label{conc}
I have shown how meson physics in presence of chemical potentials can be described using Chiral Perturbation Theory. This lead to three different phases, a normal phase, a pion condensation phase and a kaon condensation phase. The condensed phases have a charge and thus can generate mixing among mesons.\\
In this context I have illustrated how the mixing is influenced by model symmetries and that groups theory constraints the mixing possibilities. These results, obtained by group theory alone, is expected to hold in any theory describing meson states. Then I have listed the masses of mesons in the three phases, calculated using the Lagrangian (\ref{eq:Lagrangian_general}). These masses are in perfect agreement with the group theory reasoning.\\
I have also described how the charged pion decay is influenced by the chemical potentials, showing that their inclusion in the model leads to a significant asymmetry between the decay in $\ell^+\n_\ell$ and the decay in $\ell^-\bar{\n}_\ell$.\\
These results can be applied for example in the physics of compact stars, the study of the cosmic ray, the study of some nuclear decays.


\begin{thebibliography}{}
%
%

\bibitem{Migdal:1990vm}
A.~B. Migdal, E.~Saperstein, M.~Troitsky, and D.~Voskresensky, {\em Phys.Rept.}, \textbf{192}, pp.~179--437,
  1990.

\bibitem{Son:2000xc}
D.T.~Son and M.~A. Stephanov, {\em
  Phys.Rev.Lett.}, \textbf{86}, pp.~592--595, 2001.

\bibitem{Kogut:2001id}
J.~Kogut and D.~Toublan, 
  {\em Phys.Rev.}, \textbf{D64}, p.~034007, 2001.

\bibitem{Kogut:1999iv}
J.~Kogut, M.~A. Stephanov, and D.~Toublan, {\em Phys.Lett.}, \textbf{B464}, pp.~183--191, 1999.

\bibitem{Kogut:2000ek}
J.~Kogut, M.~A. Stephanov, D.~Toublan, J.~Verbaarschot, and A.~Zhitnitsky,
  {\em Nucl.Phys.},
 \textbf{B582}, pp.~477--513, 2000.

\bibitem{Son:2000by}
D.~Son and M.~A. Stephanov, {\em Phys.Atom.Nucl.}, \textbf{64},
  pp.~834--842, 2001.

\bibitem{Loewe:2002tw}
M.~Loewe and C.~Villavicencio,  {\em Phys.Rev.},\textbf{D67}, p.~074034, 2003.

\bibitem{Loewe:2004mu}
M.~Loewe and C.~Villavicencio,  {\em Phys.Rev.}, \textbf{D70}, p.~074005, 2004.

\bibitem{He:2005nk}
L.-y. He, M.~Jin, and P.-f. Zhuang, {\em Phys.Rev.},\textbf{D71}, p.~116001, 2005.

\bibitem{Xia:2014bla}
T.~Xia and P.~Zhuang,  arXiv:1411.6713v1 [hep-ph], 2014.

\bibitem{Alford:1998sd}
M.~G. Alford, A.~Kapustin, and F.~Wilczek, {\em Phys.Rev.}, \textbf{D59},
  p.~054502, 1999.

\bibitem{Kogut:2002zg}
J.~Kogut and D.~Sinclair, {\em Phys.Rev.},\textbf{D66}, p.~034505, 2002.

\bibitem{Toublan:2003tt}
D.~Toublan and J.~Kogut,  {\em
  Phys.Lett.}, \textbf{B564}, pp.~212--216, 2003.

\bibitem{Barducci:2004tt}
A.~Barducci, R.~Casalbuoni, G.~Pettini, and L.~Ravagli,  {\em Phys.Rev.},\textbf{D69}, p.~096004, 2004.

\bibitem{Barducci:2004nc}
A.~Barducci, R.~Casalbuoni, G.~Pettini, and L.~Ravagli, {\em Phys.Rev.}, \textbf{D71},
  p.~016011, 2005.

\bibitem{Klein:2004hv}
B.~Klein, D.~Toublan, and J.~Verbaarschot,  {\em Phys.Rev.},\textbf{D72},
  p.~015007, 2005.

\bibitem{Barducci:1990sv}
A.~Barducci, R.~Casalbuoni, S.~De~Curtis, R.~Gatto, and G.~Pettini, {\em Phys.Rev.},
  \textbf{D42}, pp.~1757--1763, 1990.

\bibitem{Dominguez:1993kr}
C.~Dominguez, M.~Loewe, and J.~Rojas,  {\em Phys.Lett.}, \textbf{B320}, pp.~377--380, 1994.

\bibitem{Loewe:2011tm}
M.~Loewe and C.~Villavicencio,  arXiv:1107.3859 [hep-ph], 2011.

\bibitem{Gasser:1983yg}
J.~Gasser and H.~Leutwyler,  {\em
  Annals Phys.},  \textbf{158}, p.~142, 1984.

\bibitem{Leutwyler:1993iq}
H.~Leutwyler,  {\em
  Annals Phys.}, \textbf{235}, pp.~165--203, 1994.

\bibitem{Ecker:1994gg}
G.~Ecker,  {\em Prog.Part.Nucl.Phys.},
   \textbf{35}, pp.~1--80, 1995.

\bibitem{Scherer:2002tk}
S.~Scherer, {\em
  Adv.Nucl.Phys.},  \textbf{27}, p.~277, 2003.

\bibitem{Scherer:2005ri}
S.~Scherer and M.~R. Schindler,  arXiv:hep-ph/0505265,
  2005.

\bibitem{Mammarella:2015pxa}
  A.~Mammarella and M.~Mannarelli,
  arXiv:1507.02934 [hep-ph].

%
%

\end{thebibliography}
\end{document}